\documentclass{emulateapj}
\usepackage{graphicx}
\def\Qdel{$\Delta^2_{\rm Q}$}
\def\Pdel{$\Delta^2_{\rm NL}$}
\def\LCDM{$\Lambda$CDM~}
\def\zp{z_{phot}}
\def\bzp{\bar{z}_{phot}}
\def\geqsim{\lower.73ex\hbox{$\sim$}\llap{\raise.4ex\hbox{$>$}}$\,$}
\def\leqsim{\lower.73ex\hbox{$\sim$}\llap{\raise.4ex\hbox{$<$}}$\,$}
\def\invMpch {~h~{\rm Mpc}^{-1}}
\def\Mpch {~h^{-1}~{\rm Mpc}}
\def\kpch {~h^{-1}~{\rm kpc}}

\newcommand{\myemail}{admyers@astro.uiuc.edu}
\newcommand{\Hii}{\ion{H}{2}}

\slugcomment{ApJ, in prep \today}
\shorttitle{Clustering of 300,000 Photo-Classified QSOs}
\shortauthors{Myers et al.}

\begin{document}

\title{Clustering Analyses of 300,000 Photometrically Classified Quasars--{\rm I}. Luminosity and Redshift Evolution in Quasar Bias}

\author{Adam D. Myers\altaffilmark{1,2}, Robert J. Brunner\altaffilmark{1,2}, Robert C. Nichol\altaffilmark{3}, Gordon T. Richards\altaffilmark{4,5}, Donald P. Schneider\altaffilmark{6} and Neta A. Bahcall\altaffilmark{7}}

\email{\myemail}

\altaffiltext{1}{Department of Astronomy, University of Illinois at Urbana-Champaign, Urbana, IL 61801}
\altaffiltext{2}{National Center for Supercomputing Applications, Champaign, IL 61820}
\altaffiltext{3}{ICG, Mercantile House, Hampshire Terrace, University of Portsmouth, Portsmouth, P01 2EG, UK}
\altaffiltext{4}{Department of Physics \& Astronomy, The Johns Hopkins University, 3400 N Charles St, Baltimore, MD 21218}
\altaffiltext{5}{Department of Physics, Drexel University, 3141 Chestnut Street, Philadelphia, PA 19104}
\altaffiltext{6}{Department of Astronomy and Astrophysics, 525 Davey Laboratory, Pennsylvania State University, University Park, PA 16802}
\altaffiltext{7}{Department of Astrophysical Sciences, Princeton University, Princeton, NJ 08544}

\begin{abstract}
Using $\sim$300,000 photometrically classified quasars, by far the largest quasar sample ever used for such analyses, we study the redshift and luminosity evolution of quasar clustering on scales of $\sim$50$\kpch$ to $\sim$20$\Mpch$ from redshifts of $\bar{z}\sim$0.75 to $\bar{z}\sim$2.28. We parameterize our clustering amplitudes using realistic dark matter models, and find that a \LCDM power spectrum provides a superb fit to our data with a redshift-averaged quasar bias of $b_Q^{\bar{z}=1.40} = 2.41\pm0.08$ ($P_{<\chi^2}=0.847$) for $\sigma_8=0.9$. This represents a better fit than the best-fit power-law model ($\omega = 0.0493\pm0.0064\theta^{-0.928\pm0.055}$; $P_{<\chi^2}=0.482$). We find $b_Q$ increases with redshift. This evolution is significant at $>99.6$\% using our data set alone, increasing to $>99.9999$\% if stellar contamination is not explicitly parameterized. We measure the quasar classification efficiency across our full sample as $a = 95.6 \pm ^{4.4}_{1.9}$\%, a star-quasar separation comparable with the star-galaxy separation in many photometric studies of galaxy clustering. We derive the mean mass of the dark matter halos hosting quasars as $M_{DMH}=5.2\pm0.6\times10^{12}~{\rm h^{-1}}M_\sun$. At $\bar{z}\sim1.9$ we find a $1.5\sigma$ deviation from luminosity-independent quasar clustering; this suggests that increasing our sample size by a factor of $\sim$1.8 could begin to constrain any luminosity dependence in quasar bias at $z\sim2$. Our results agree with recent studies of quasar environments at $z < 0.4$, which detected little luminosity dependence to quasar clustering on proper scales $\geqsim~50\kpch$. At $z < 1.6$, our analysis suggests that $b_Q$ is constant with luminosity to within $\Delta b_Q\sim0.6$, and that, for $g < 21$, angular quasar autocorrelation measurements are unlikely to have sufficient statistical power at $z~\leqsim~1.6$ to detect any luminosity dependence in quasars' clustering.
\end{abstract}


\section{Introduction}

As the form of the nonbaryonic, cold dark matter that underpins mass in the cosmos becomes increasingly accurately described (e.g., \citealt{Col05}), understanding the baryonic processes that fuel quasars and trigger galaxy formation becomes an increasingly realistic endeavor. It is now established that most, if not all, local galaxies harbor a supermassive black hole \citep{Kor95,Ric98} and that the mass of these black holes correlate with several properties of their host galaxy's bulge \citep{Mag98,Fer00,Geb00,Gra02,Tre02,Wyi06}, implying a causal link between black holes and star formation in galaxy spheroids (e.g., \citealt{Sil98}). Observational evidence is accumulating to suggest that such a causal link remains at higher redshift \citep{Shi03}.

It has long been suspected that accretion of baryons onto supermassive black holes is responsible for the powerful UV-excess (UVX) emission seen in quasars (see, e.g., \citealt{Ree84} for a review), so the role of supermassive black holes in galaxy formation suggests an interplay between nascent galaxies and quasar activity. A symbiotic view of galaxy formation has emerged, in which galaxy mergers drive the formation of quasars, and supermassive black holes, which in turn seed new galaxies (e.g., \citealt{Hec86,Car90,Bar92,Lac93,DiM05}). Given that only merging systems of a certain minimum mass can trigger a UVX quasar phase visible against background star formation in the host galaxy (e.g., \citealt{Hop06}), this picture is consistent with emerging evidence that at $z~\leqsim~2.5$ quasar bias evolves with redshift but that quasars inhabit dark matter halos of similar average mass at every redshift \citep{Por04,Cro05,Mye06}. However, the simplicity of this picture belies a rich complexity in the important physical processes that entwine quasar, galaxy and star formation (see \citealt{Hop06} for a review). Some of the many theoretical insights into this complexity have included the importance of galaxy mergers (e.g., \citealt{Too72,Whi79,Neg82,Bar92,Fra99}), cooling flows (e.g., \citealt{Cio97,Cio01}), heating through various feedback mechanisms (e.g., \citealt{Sil98,Wyi02,Spr03}) and/or the eventual cutoff of accretion onto a central black hole by gas ejection (e.g., \citealt{Sil98,Fab99,Saz05}).

Clearly, quasar evolution is an important tracer of galaxy formation; however, the large number of components that help regulate models of quasar activity beg new constraints.  Measurements of quasar clustering amplitudes, which directly correlate with the average mass of the halos that harbor quasars, have provided useful broad constraints on gravitationally driven aspects of quasar evolution.  However, the relevance of gas physics to quasar evolution means that measurements of the luminosity function of quasars (see \citealt{Ric06} for a review) have also proved key in constraining baryonic elements of quasar evolution, e.g., quasar lifetimes via the ``duty cycle" \citep{Hai01,Mar01}. \citet{Hop05b} have suggested that the peak luminosity distribution of quasars is fundamental to characterizing the quasar population; this suggests that important observational constraints will emerge by considering baryons and gravity in tandem, by measuring the luminosity evolution of quasar clustering (e.g., \citealt{Val01,Lid06}).

Since the first significant detections of quasar clustering, the two-point correlation function (e.g., \citealt{Tot69,Pee80}) has frequently been used to measure the amplitude of quasar clustering, and accuracy has improved in step with sample sizes.  Recent detections, in the wake of large spectroscopic quasar surveys, have led to some confidence about the evolution of quasar clustering (e.g., \citealt{Cro05}; henceforth Cro05), constraining a lack of evolution in the dark matter halos that host quasars to $\sim$50\%.  However, the dependence of quasar clustering on luminosity appears to be quite weak (e.g., Cro05; \citealt{Lid06}), and probing variations in quasar clustering as a function of luminosity or other physical properties, as well as more tightly constraining evolution in quasar clustering, is limited by quasar sample sizes.

Large samples of photometrically selected objects have long been used to probe angular galaxy clustering \citep{Gro77}, leading to important cosmological constraints such as early detections \citep{Mad90a} of deviations from Standard Cold Dark Matter (CDM) models or early detections of Dark Energy (e.g., \citealt{Scr03}). Such angular analyses were complimentary to analyses using spectroscopic data because of the larger numbers of galaxies that could be photometrically selected. Similar angular clustering analyses of complimentarily large samples of photometrically selected quasars were impossible, however; star-galaxy separation became efficient in galaxy surveys with the advent of automatic plate measurements (e.g., 90-95\% efficient in the APM survey; \citealt{Mad90b}) but star-quasar separation lagged behind (e.g.,$\sim$60\% efficient in the 2dF QSO Redshift Survey; \citealt{Cro04}; henceforth 2QZ).  With the recent advent of sophisticated photometric classification of quasars \citep{Ric04} star-quasar separation at many redshifts is now highly efficient ($\sim$95\%; \citealt{Ric04,Mye06}), meaning that large quasar samples can be used to measure angular quasar clustering as a function of physical properties \citep{Mye06}, and to use angular quasar clustering to probe Dark Matter \citep{Scr05} and Dark Energy \citep{Gia06}.

In \citeauthor{Mye06} (\citeyear{Mye06}; henceforth Mye06), we presented a first, proof-of-concept, analysis of the clustering of $\sim$80,000 photometrically classified quasars.  In this series of papers, we extend this work, improving our modeling techniques and presenting measurements of the two-point correlation function of $\sim$300,000 photometrically classified quasars drawn from the fourth data release (DR4) of the Sloan Digital Sky Survey (SDSS; e.g., \citealt{Sto02,Aba03,Aba04,Aba05}).  Our goal in this paper is to study the dependence of quasar clustering on redshift and luminosity, focusing on linear and quasi-linear scales. In a companion paper (\citealt{Mye07}; henceforth Paper2), we analyze quasar clustering on smaller scales. 

Extensive details of our techniques, modeling, and systematics are presented in Appendixes A and B, allowing our main analysis to be presented in Section~\ref{sec:quasclus}, after detailing our data sample in Section~\ref{sec:data}. Our main results are ordered in a concluding section (Section~\ref{sec:summ}). Unless otherwise specified, we assume a \LCDM cosmology with ($\Omega_m$, $\Omega_\Lambda$, $\sigma_8$, $\Gamma$, $h\equiv H_0/100{\rm km~s^{-1}~Mpc^{-1}}$)$ = (0.3,0.7,0.9,0.21,0.7)$, where $\Gamma$ is the shape of the matter power spectrum ($\Gamma=\Omega_m h$ for baryon-free CDM). We correct all magnitudes for Galactic extinction using the dust maps of \citet{Sch98}.

\section{The DR4 KDE Sample}
\label{sec:data}

The quasar sample that we analyze is constructed using the Kernel Density Estimation (KDE) technique of \citet{Ric04}, which draws on many unique technical aspects of the SDSS (e.g., \citealt{Yor00}), including superior photometry (e.g., \citealt{Fuk96,Gun98,Lup99,Hog01,Smi02,Ive04}), astrometry (e.g., \citealt{Pie03}), and data acquisition (e.g., \citealt{Gun06,Tuc06}). As in \citet{Ric04}, the sample is restricted to SDSS point sources with $u-g < 1$, (observed) $g \geq 14.5$ and (dereddened) $g < 21$. Separations, in 4-dimensional color-space, from a sample of $\sim$10\% of point sources in SDSS Data Release 1 (DR1; \citealt{Aba03}) and from the quasar sample of \citeauthor{Sch03} (\citeyear{Sch03}; henceforth DR1QSO) are determined for each object to be classified. A Bayesian classifier then assigns each object a probability of being a ``quasar" or ``star". Taken in logarithmic ratio, the distribution of these probabilities is sufficiently bimodal to separate $z~\leqsim~2.5$ quasars from stars with $\sim$95\% efficiency (\citealt{Ric04}; see also Mye06). Applying the KDE technique to SDSS DR4, results in our DR4 KDE sample of 344,431 objects\footnote{Available at http://sdss.ncsa.uiuc.edu/qso/nbckde}, a sample $3.5\times$ larger than the DR1 sample used in Mye06. Each KDE object is assigned a photometric redshift estimate as described in \citet{Wei04}.

To meaningfully model quasar clustering we must know the normalized redshift distribution of our sources (${\rm d}N/{\rm d}z$ in Equation~\ref{eqn:projmod}). For consistency with Mye06 (see their Figure~6), we estimate ${\rm d}N/{\rm d}z$ from spectroscopic matches to DR1QSO, matches to spectra taken from the SDSS second data release (\citealt{Aba04}; henceforth DR2) or matches to the 2QZ. As our KDE technique is currently trained on DR1QSO, obtaining ${\rm d}N/{\rm d}z$ from DR1QSO is arguably a fairer approach than using quasars from later data releases. Mye06 demonstrated that including or excluding matches with the 2QZ or DR2 has little affect on the form of ${\rm d}N/{\rm d}z$, and further showed that the methodology of using spectroscopic matches is broadly consistent with estimating ${\rm d}N/{\rm d}z$ from photometric redshifts \citep{Wei04}.

\begin{figure}
\plotone{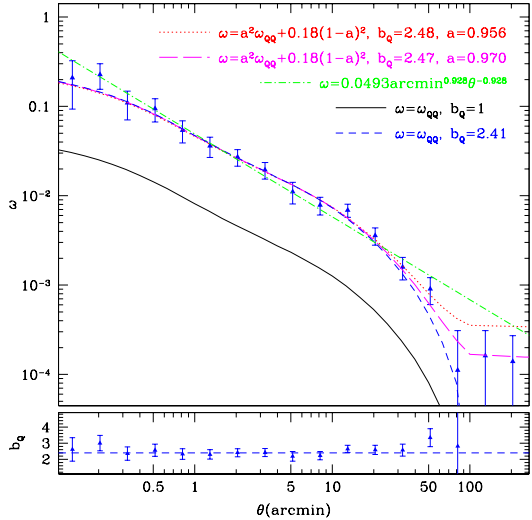}
\caption{The autocorrelation of all 299,276 ($A_g < 0.21$) DR4 KDE objects in our working area. Errors are jackknifed with a resolution of $10^\circ$. The short-dashed line is the best fitting bias model (Equation~\ref{eqn:projmod}) over the range 0.16$'$ to 63$'$ ($\sim$55$\kpch$ to $22\Mpch$ at the DR4 KDE sample's mean redshift of $z\sim1.4$). The dotted line is the best fitting bias model (over the same range) when fitting for stellar contamination in the KDE sample (Equation~\ref{eqn:stelcon} with $\omega_{SS}=0.18$). The long-dashed line is the best fitting bias model with stellar contamination when the fitting range is extended out to 100$'$ ($\sim$35$\Mpch$). The solid line shows the expected clustering for a linear bias model, which is easily rejected at extremely high significance. The lower panel shows the measured quasar bias relative to the linear bias model. The dot-dashed line is the best-fit power-law over 0.16$'$ to 63$'$. We estimate ${\rm d}N/{\rm d}z$ for the redshift distribution of the full DR4 KDE sample using a simple spline fit.\label{fig:allKDE}}
\end{figure}

\section{Quasar Clustering Results}
\label{sec:quasclus}

\subsection{Mean Quasar Bias at $z\sim1.4$}
\label{sec:fullcorr}

In Figure~\ref{fig:allKDE} we show the ($A_g < 0.21$) DR4 KDE autocorrelation. We fit bias models (see Equation~\ref{eqn:projmod}) over scales of 0.16$'$ to 63$'$ ($\sim$55$\kpch$ to $22\Mpch$ at the DR4 KDE sample's mean redshift of $\bar{z}=1.4$). The fit's upper scale limit is nominally set by stellar contamination (see Appendix~\ref{sec:stelcon}) and the lower limit is set by the dark matter model we use for $\omega_{QQ}$. Smi03 note that their models accurately reproduce $\Delta_{\rm NL}^2$ to the limits of current simulations ($\sim$3\%) at wavenumbers of $k  < 10\invMpch$ ($\geqsim~1'$); however, their models appear quite accurate even on scales several times smaller than this (see, e.g., Figure~15 of Smi03) and, in any case, their models remain useful as a phenomenological description of dark matter clustering on all our scales of interest.  In particular, any models that augment the approach of Smi03 at $k < 10\invMpch$ should be easy to compare to Smi03, and thus to our results.

Our best fit bias model to the DR4 KDE autocorrelation has $b_Q^{\bar{z}=1.40} = 2.41\pm0.08$ ($P_{<\chi^2}=0.847$), in good agreement with Mye06 ($b_Q^{\bar{z}=1.40} = 2.51\pm0.46$) but with considerably more precision. This is preferred over the best-fit power-law model ($\omega = 0.0493\pm0.0064\theta^{-0.928\pm0.055}$; $P_{<\chi^2}=0.482$). A linear bias model (see Figure~\ref{fig:allKDE}), is ruled out at an extremely high level of significance ($P_{<\chi^2}\sim10^{-23}$). Our measured $b_Q$ is in reasonable agreement with the value of $b_Q^{\bar{z}=1.47} = 2.42\pm^{0.20}_{0.21}$ obtained by \citeauthor{Por04} (\citeyear{Por04}; henceforth PMN04) in a clustering analysis of 14,000 $M_{b_J} < -22.5$ 2QZ quasars, even though PMN04 use $\sigma_8=0.8$, which will inflate their result by $\sim$13\% as compared to our use of $\sigma_8=0.9$. Our result is slightly at odds with the value of $b_Q^{\bar{z}=1.35} = 2.02\pm0.07$ found by Cro05 for the full 2QZ sample after correcting their result for redshift-space distortions. It is, however, well within the error bars of their empirical fit to the evolution of quasar clustering of $b_Q = (0.53\pm0.19)+(0.289\pm0.035)(1+z)^2$. The form of this empirical fit broadly implies that our result (at $\bar{z}=1.40$) should be $\sim5$\% lower than an estimate at $\bar{z}=1.47$ and $\sim3$\% higher than at $\bar{z}=1.35$.

To test stellar contamination effects, we also fit a model of the form $a^2\omega_{QQ} + (1-a)^2\omega_{SS}$, where $a$ is the KDE efficiency and $\omega_{SS}=0.18$ (as derived in Mye06) and find $b_Q=2.48\pm0.15$ and $a = 0.956 \pm ^{0.044}_{0.019}$  ($P_{<\chi^2}=0.920$). As we previously argued in Mye06, stellar contamination causes measurable deviation from the true $\omega_{QQ}$ only on scales of a degree or more. An appropriate approach would be to determine $a$ at $> 1^\circ$ and input this value to fits on scales $<1^\circ$. If $a~\geqsim~0.9$, this largely becomes unnecessary, as introducing an efficiency term does not significantly alter measured values of $b_Q$ at $<1^\circ$. 

In Figure~\ref{fig:allKDE}, we also fit a stellar contamination model to larger scales (where stellar contamination begins to dominate fits) as such a model should still be valid on these scales. When a stellar contamination model is fit out to 100$'$, we obtain $b_Q=2.47\pm0.15$ and $a = 0.970 \pm ^{0.030}_{0.017}$  ($P_{<\chi^2}=0.908$). This value of $a$ reproduces the data very well out to at least $7^\circ$ (see Figure~\ref{fig:allKDE}); however, although altering the scale of the fit affects estimates of $a$, $b_Q$ is essentially unchanged, as stellar contamination only dominates on large scales. We further note that altering $\omega_{SS}$ to 0.25, as is appropriate at $\sim$30$'$ (see Mye06) barely changes our result, giving $b_Q=2.47\pm0.14$ and $a = 0.974 \pm ^{0.026}_{0.015}$  ($P_{<\chi^2}=0.936$); again, simply because stellar contamination is only influential on scales larger than those we typically fit. In general, throughout this paper, we will quote results for models that simultaneously fit for $a$ and $b_Q$.  However, in most cases fitting for $a$ at $<1^\circ$, although illustrative, is overkill, and falsely reduces the significance of our $b_Q$ measurements. We will therefore trust the significance estimates of those fits that ignore stellar contamination.
 
 \begin{figure}
\plotone{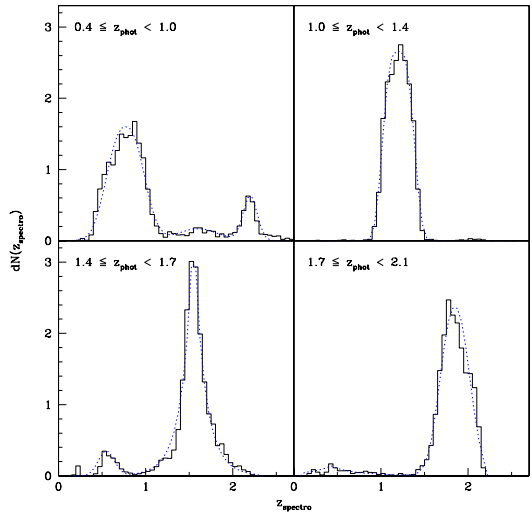}
\caption{The spectroscopic redshift ($z_{spectro}$) distribution of KDE quasars for some photometric redshift bins ($z_{phot}$) used in our analyses. The dotted lines are functional fits of the form described by Equation~\ref{eqn:multgauss}. The plotted distributions have been normalized and the integral under the fitted functions is always within 0.1\% of unity\label{fig:zbinhisto}}
\end{figure}

\subsection{Evolution in Quasar Bias}
\label{sec:evo}

The spectroscopic redshift distribution of quasars at a given photometric redshift can become increasingly complex as the photometric redshift bandwidth is reduced. Therefore, to model the evolution of quasar clustering in a number of photometric redshift bins, it is convenient to have a better mechanism for modeling ${\rm d}N/{\rm d}z$ than the simple spline fit used in Section~\ref{sec:fullcorr}. Typically, the photometric redshift distributions we study have a primary peak where the photometric solution agrees with the spectroscopy and, in some cases, a minor, secondary peak where the photometric solution is inaccurate (due to so-called {\it catastrophic failures}). We therefore adopt the approach of summing a number of functions of the form

\begin{equation}
{\rm d}N = \sum_i\beta_i\exp\frac{-\left|z-\bar{z_i}\right|^{n_i}}{n_i\sigma_i^{n_i}}{\rm d}z
\label{eqn:multgauss}
\end{equation}

\noindent where $n$ (typically close to the Gaussian value of 2), $\sigma,\bar{z}$ and $\beta$ are free parameters. These functions can excellently reproduce ${\rm d}N/{\rm d}z$ (see Figure~\ref{fig:zbinhisto}). Our schema for binning in photometric redshift is chosen to maximize object numbers in each bin while limiting discrepancies between the photometric and spectroscopic redshift estimates (see Mye06).

\begin{figure}
\plotone{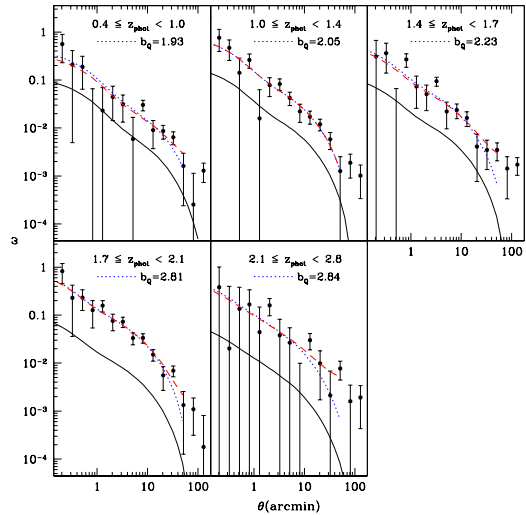}
\caption{Quasar clustering evolution as a function of photometric redshift. There are $\sim$65,000 quasars in each bin except for the $2.1 \leq z_{phot} < 2.8$ bin, which contains $\sim$28,000 quasars.
The solid line is the expected clustering of dark matter derived from Smi03. The dotted line is our best fit  model where only the quasar bias, $b_Q$, is varied. The dashed line is a two-parameter model that incorporates stellar contamination as well as quasar bias (see Table~\ref{table:1} for model values). Errors in this plot are jackknifed and fits are made over scales of 0.16$'$ to 63$'$ using a full covariance matrix. A scale of 0.16$'$ to 63$'$ is $\sim$55$\kpch$ to $22\Mpch$ in all bins except the lowest redshift bin, where the scales are slightly reduced to $\sim$50$\kpch$ to $20\Mpch$.
\label{fig:zbins}}
\end{figure}

\begin{figure}
\plotone{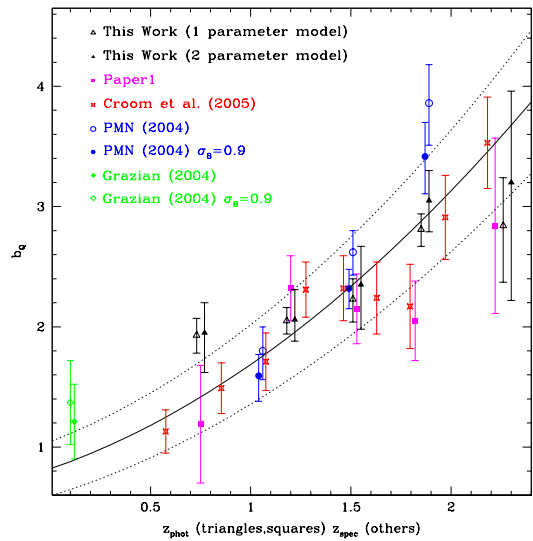}
\caption{Our estimates of the evolution of quasar bias compared to other authors, and to our estimates from Mye06. The solid line plots the semi-empirical relationship $b_Q(z) = 0.53+0.289(1+z)^2$ derived by Cro05, and the dotted lines track the 1$\sigma$ error. The open triangles show our best fit model where only the quasar bias, $b_Q$, is varied. The solid triangles show the results when the stellar contamination, is also allowed to vary (see Table~\ref{table:1} for model values). PMN04 and \citet{Gra04} use $\sigma_8=0.8$, so we also show the effect of projecting their results to match our chosen value of $\sigma_8=0.9$. In Section~\ref{sec:fullcorr}, we note that, due to the photometric nature of the redshifts we use, our points at the lowest and highest redshift are almost certainly {\it at least} 10\% too high and 10\% too low, respectively. Points at other redshifts are not at all biased by the use of photometric redshifts.\label{fig:bias}}
\end{figure}

In Figure~\ref{fig:zbins} we show the evolution of the angular quasar autocorrelation with photometric redshift. We derive estimates of the quasar bias using Equation~\ref{eqn:projmod}, and also consider a two-parameter stellar-contamination model (Equation~\ref{eqn:stelcon}). In Figure~\ref{fig:bias} we display our measured quasar bias evolution. Our data alone (i.e., without assuming, $b_Q\sim1$ at $z=0$), rules out constant $b_Q$ at all redshifts at $>99.99$\%, dropping to $>97.9$\% if stellar contamination is allowed to freely vary. As discussed in Section~\ref{sec:fullcorr}, not fitting for stellar contamination likely better estimates significances. Using 2QZ data, PMN04 and Cro05 have independently determined that $b_Q$ evolves with redshift. We note that, after correcting for differing $\sigma_8$, our value of $b_Q^{\bzp=1.87}$ disagrees with $b_Q^{z=1.89}$ from PMN04 but only at the 1.8$\sigma$ level, dropping to 0.8$\sigma$ if stellar contamination is incorporated. The $M_{b_J} < -22.5$ restriction adopted by PMN04 is immaterial in this context, as every $g < 21$ quasar at $z > 1.7$ should be brighter than $M_{b_J} = -22.5$. Bias determinations derived from cosmological models that are more like those used by PMN04 and Cro05 are provided in~Table~\ref{table:1} (see also the discussion in Section~\ref{sec:DMH}).

The photometric nature of our redshifts leads to problems for our $\zp=0.75$ and $\zp=2.28$ bins (i.e., our lowest and highest redshift bins); in particular, where to plot these bins on the $\zp$ axis of Figure~\ref{fig:bias}. At $1 < \zp < 2$, the mean redshift of our spectroscopic matches is close to $\bzp$, but beyond this range the true redshift is further from the photometric estimate, due to catastrophic failures in the photometric redshift estimation. Figure~\ref{fig:zbinhisto} demonstrates that at $\zp < 1$ ($\zp > 2$) there is a secondary solution, amounting to 12.6\% (21.2\%) of the area under ${\rm d}N/{\rm d}z$, at $\zp > 2$ ($\zp < 1$). We can examine clustering at these secondary $\zp$ solutions by adopting a similar approach to Equation~\ref{eqn:stelcon}:

\begin{equation}
B^2\omega_{1} =  \omega_{1+2}-(1-B)^2\omega_{2}
\end{equation}

\noindent Where, $\omega_{1}$, $\omega_{2}$ and $\omega_{1+2}$ are, respectively, the true clustering at the primary and secondary $\zp$ solutions, and the clustering we measure as a combination of the two $\zp$ solutions. $B$ is the relative contribution of the primary and secondary $\zp$ solutions to ${\rm d}N/{\rm d}z$. We can then estimate the true value of the bias at the position of the primary solution in $\zp$ as 

\begin{equation}
b_1^2 = \frac{\omega_1}{\omega_1^{M}} = \frac{b_{1+2}^2\omega_{1+2}^{M}-(1-B)^2b_{2}^2\omega_{2}^{M}}{B^2\omega_{1}^{M}} 
\label{eqn:reweight}
\end{equation}

\noindent where the superscript $M$ denotes the model values of $\omega$ (Equation~\ref{eqn:projmod} with $b_Q=1$), and the $b_i$ denote quasar bias. Although we don't know the true values of $b_1$ and $b_2$, they can be estimated from measurements of $b_Q$.

If we follow this analysis, our value of $b_Q$ at $\bzp=0.75$ ($\bzp=2.28$) should be {\it reduced} ({\it increased}) by {\it at least} 10\%. The true values of $b_Q$ must lie beyond even these 10\% offsets, because we must use our measured values of $b_Q$ at the secondary solution in $\zp$, rather than the (unknown) true value, to estimate the true value in the primary $\zp$ bin. If we lower our estimates of $b_Q^{\bzp=0.75}$ by 10\% and increase our estimates of $b_Q^{\bzp=2.28}$ by 10\%, we find that our data rule out a constant $b_Q$ at all redshifts with a significance of $>99.9999$\%, dropping to $>99.6$\% if stellar contamination is allowed to freely vary. We note that applying Equation~\ref{eqn:reweight} to our bins at $\bzp=1.20,1.53,1.87$ has no affect on $b_Q$ values, as any secondary solutions in $\zp$ have a very small weight ($B~\leqsim~0.05$).

\begin{figure}
\plotone{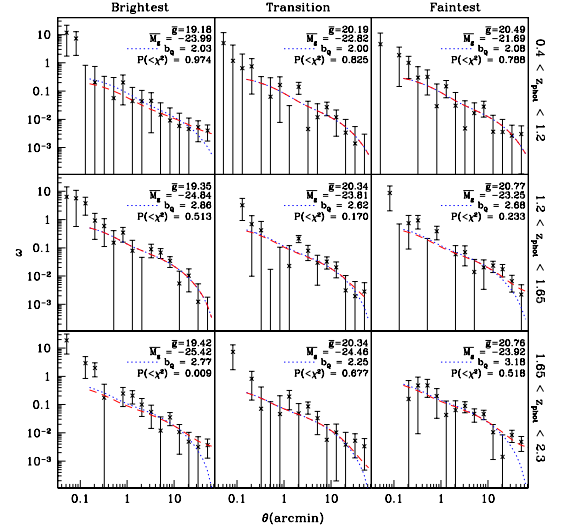}
\caption{Quasar clustering evolution as a bivariate function of absolute magnitude, $M_g$, and photometric redshift, $\zp$. The rows of the nine panels show three bins of roughly equal numbers in $\zp$. The columns divide each $\zp$ bin into three of equal numbers in $M_g$. There are $\sim$30,000 quasars in each bin. Labeled in each panel are the mean $g$ apparent and absolute magnitudes, the quasar bias, $b_Q$, derived as in Figure~\ref{fig:zbins}, and the $\chi^2$ probability of our best fit model where only $b_Q$ is fitted (the dotted line). The dashed line is a two-parameter model that incorporates stellar contamination as well as quasar bias. Table~\ref{table:3} displays model values. Errors in this plot are jackknifed and fits are made over scales of 0.16$'$ to 63$'$ using a full covariance matrix. A scale of 0.16$'$ to 63$'$ is $\sim$55$\kpch$ to $22\Mpch$ in all bins except the lowest redshift bin, where the scales are slightly reduced to $\sim$50$\kpch$ to $20\Mpch$.\label{fig:absmag}}

\end{figure}
\begin{figure}
\plotone{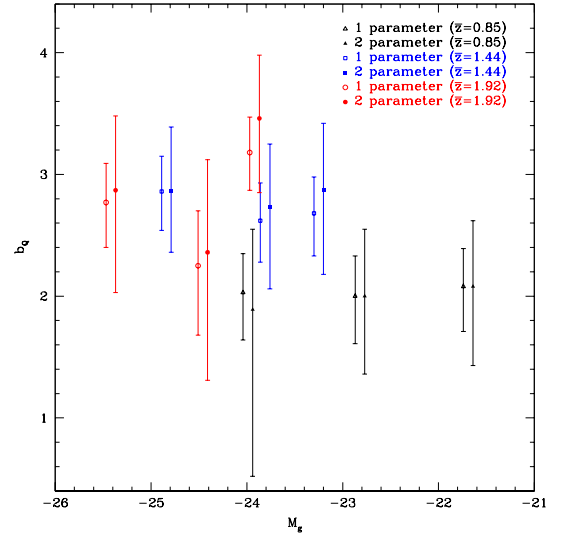}
\caption{Our quasar bias estimates as a bivariate function of $g$-band absolute magnitude, $M_g$, and photometric redshift, $\zp$. The mean photometric redshift that corresponds to each shape is labeled in the plot. The open shapes show our best fit model when only the quasar bias, $b_Q$, is varied. The solid shapes show estimates when a second parameter, the stellar contamination, is also allowed to vary (see Table~\ref{table:3} for model values).\label{fig:biasabsmag}}
\end{figure}

\begin{deluxetable*}{ccccccccccc}
\tabletypesize{\scriptsize}
\tablecaption{Estimates of the quasar bias, $b_Q$ and the quasar host halo mass $M_{DMH}$ as a function of photometric redshift $\zp$\label{table:1}}
\tablecolumns{11}
\tablewidth{0pt}
\tablehead{
\colhead{}&\colhead{} & \colhead{} & \multicolumn{5}{c}{$\Gamma=0.21,\sigma_8=0.9$} & \colhead{} &\multicolumn{2}{c}{$\Gamma=0.15,\sigma_8=0.8$}  \\ \cline{4-8}\cline{10-11} \\
 \multicolumn{2}{c}{$\zp$} & \colhead{} & \multicolumn{2}{c}{1 parameter model} & \colhead{} & \multicolumn{2}{c}{2 parameter model\tablenotemark{a}} & \colhead{} & \multicolumn{2}{c}{1 parameter model}\\ \cline{1-2}\cline{4-5}\cline{7-8}\cline{10-11} \\
 \colhead{range} & \colhead{mean} & \colhead{} & \colhead{$b_Q$} & \colhead{$M_{DMH} ({\rm h^{-1}}M_\sun)$}&\colhead{} & \colhead{$a$} & \colhead{$b_Q$} & \colhead{} & \colhead{$b_Q$} &\colhead{$M_{DMH} ({\rm h^{-1}}M_\sun)$}
}
\startdata
$0.4, 2.3$ & 1.40 && $2.41\pm^{0.08}_{0.09}$ &$12.1\pm^{1.3}_{1.4}\times10^{12}$&& $0.956\pm^{0.044}_{0.019}$ & $2.48\pm^{0.13}_{0.15}$ &&$2.57\pm^{0.10}_{0.09}$&$5.72\pm^{0.84}_{0.70}\times10^{12}$\\
$0.4, 1.0$\tablenotemark{c} & 0.75 && $1.93\pm^{0.14}_{0.15}$ &$8.90\pm^{2.71}_{2.46}\times10^{12}$&& $0.903\pm^{0.097}_{0.041}$ & $1.95\pm^{0.25}_{0.33}$ &&$2.06\pm^{0.14}_{0.16}$&$4.08\pm^{1.32}_{1.25}\times10^{12}$\\
$1.0, 1.4$ & 1.20 && $2.05\pm^{0.11}_{0.11}$ &$12.5\pm^{2.3}_{2.1}\times10^{12}$&& $0.997\pm^{0.003}_{0.092}$ & $2.06\pm^{0.25}_{0.18}$ &&$2.16\pm^{0.11}_{0.12}$&$5.70\pm^{1.15}_{1.11}\times10^{12}$\\
$1.4, 1.7$ & 1.53 && $2.23\pm^{0.17}_{0.19}$ &$9.22\pm^{2.49}_{2.40}\times10^{12}$&& $0.884\pm^{0.057}_{0.036}$ & $2.35\pm^{0.32}_{0.37}$ &&$2.39\pm^{0.18}_{0.20}$&$4.30\pm^{1.33}_{1.23}\times10^{12}$\\
$1.7, 2.1$ & 1.87 && $2.81\pm^{0.13}_{0.14}$ &$11.3\pm^{1.7}_{1.7}\times10^{12}$&& $0.907\pm^{0.077}_{0.037}$ & $3.05\pm^{0.25}_{0.26}$ &&$3.00\pm^{0.14}_{0.15}$&$5.32\pm^{0.94}_{0.90}\times10^{12}$\\
$2.1, 2.8$\tablenotemark{c} & 2.28 && $2.84\pm^{0.40}_{0.47}$ &$11.3\pm^{5.0}_{4.6}\times10^{12}$&& $0.840\pm^{0.091}_{0.039}$ & $3.20\pm^{0.76}_{0.98}$ &&$3.13\pm^{0.43}_{0.50}$&$5.95\pm^{2.92}_{2.56}\times10^{12}$\\
\enddata

\tablenotetext{a}{With stellar contamination $1-a$.}
\tablenotetext{b}{Due to catastrophic failures, values of $b_Q$ in this row need lowered at least 10\%. The derived $M_{DMH}$ use the reduced values. See Equation~\ref{eqn:reweight} and the associated discussion.}
\tablenotetext{c}{Due to catastrophic failures, values of $b_Q$ in this row need raised at least 10\%. The derived $M_{DMH}$ use the increased values. See Equation~\ref{eqn:reweight} and the associated discussion.}

\end{deluxetable*}

\subsection{The Luminosity Evolution of Quasar Clustering}
\label{sec:lumclus}

While the luminosity of UVX quasars depends somewhat on the mass of the underlying black hole, the mechanisms that drive baryons onto the accretion disk feeding the black hole are also important. As such, models of quasar formation and evolution (e.g., \citeauthor{Hop05a}~\citeyear{Hop05a,Hop05b,Hop06}) can be degenerate between mass and luminosity. It is therefore useful to examine constraints on quasar bias as a function of luminosity. As discussed in Mye06, ideal tests of quasar evolution would attempt to break luminosity-redshift degeneracy and examine multivariate quasar properties. We now repeat our clustering analysis as a bivariate function of redshift and luminosity. We derive $g$-band absolute magnitudes ($M_g$) for KDE objects by assuming that each photoz is a reasonable ensemble estimate of redshift. We incorporate the K-correction from \citeauthor{Wis00} (\citeyear{Wis00}; see, e.g., Mye06). Consistent $M_g$ binning at every redshift is impractical for quasars (which span $\sim$8 magnitudes in $M_g$), so, as in Mye06, we split the KDE sample into three photometric redshift bins, then subdivide these into three $M_g$ bins. We then measure the autocorrelation of each of these nine subsamples.

In Figure~\ref{fig:absmag} we plot the bivariate quasar autocorrelation as a function of photometric redshift and absolute magnitude. As before, we derive the quasar bias over the range 0.16$'$ to 63$'$, plotting the results in Figure~\ref{fig:biasabsmag} (see also Table~\ref{table:3}). Our results are at least twice as precise as the equivalent results from Mye06, but other than the general increase with redshift discussed in Section~\ref{sec:evo}, there appears to be no discernible trend in quasar clustering with absolute magnitude. However, a model with constant quasar bias as a function of absolute magnitude is only just accepted by our data ($b_Q=2.50\pm0.11, P_{<\chi^2}=0.092$). If we instead take $b_Q$ values from models that allow a stellar contamination component, a constant $b_Q$ model is more acceptable ($b_Q=2.55\pm0.21, P_{<\chi^2}=0.662$). 

To test for luminosity-dependent bias we combine measurements for each set of two magnitudes within each redshift bin plotted in Figure~\ref{fig:biasabsmag} into a single inverse-variance-weighted estimate and determine whether this estimate would be rejected by the measurement in the third magnitude bin. Quoting the maximum rejection in each case, we find: (1) Incorporating stellar contamination, a maximum rejection of $0.2\sigma$ for $\bar{z}=0.85$, $0.2\sigma$ for $\bar{z}=1.44$ and $1.0\sigma$ for $\bar{z}=1.92$, and; (2) ignoring stellar contamination, a maximum rejection of $0.1\sigma$ for $\bar{z}=0.85$, $0.5\sigma$ for $\bar{z}=1.44$ and $1.5\sigma$ for $\bar{z}=1.92$.

\section{Discussion}

\subsection{Quasar Host Masses}
\label{sec:DMH}

Following Cro05 we can use the ellipsoidal collapse model of \citet{She01} to convert quasar biases into masses for the halos hosting UVX quasars. We weight this model across our (normalized) redshift distributions

\begin{eqnarray}
b_Q(M_{DMH},\bar{z}) = 1+\int^{z=z_{max}}_{z=z_{min}}{\rm d}z\frac{{\rm d}N}{{\rm d}z}\frac{1}{\sqrt{a}\delta_{sc}(z)}~~~~~~~~~~~~&& \\
\times \left[\sqrt{a}(a\nu^2) + \sqrt{a}b(a\nu^2)^{1-c}-\frac{(a\nu^{2})^c}{(a\nu^2)^c+b(1-c)(1-c/2)}\right] \nonumber
\end{eqnarray}

\noindent where $a=0.707$, $b=0.5$, $c = 0.6$ and $\delta_{sc}(z)$, the critical density contrast for spherical collapse, is given by \citet{Nav97} as $0.15(12\pi)^{2/3}\Omega_{mz}^{0.0055}$ (for flat cosmologies), where $\Omega_{mz}\equiv\Omega_m(z)$ is given, e.g., in Mye06.

Masses can be derived via $\nu=\delta_{sc}(z)/\sigma_r(M,z)$, where $\sigma_r(M,z)=$$\sigma_r(M)D(z)$, and $D(z)$ is the linear growth factor, which we approximate using the formula of \citeauthor{Car92} (\citeyear{Car92}; see, e.g., Mye06). The mass variance for a halo, $\sigma_r^2(M_{DMH})$ can be determined from the radius of a halo of mean mass $M_{DMH}$

\begin{equation}
r = \left(\frac{3M_{DMH}}{4\pi\rho_0}\right)^{1/3}
\end{equation}

\noindent where $\rho_0 = 2.78\times10^{11}\Omega_mh^2M_\sun{\rm Mpc^{-3}}$ is the present mean density of the universe. This mass scale implies a mass variance of

\begin{equation}
\sigma^2_r(M_{DMH}) = \frac{V}{2\pi^2}\int_0^\infty k^3 P(k)\left[\frac{3j_1(kr)}{kr}\right]^2\frac{{\rm d}k}{k}
\end{equation}

\noindent where the term in square brackets represents a spherical top hat smoothing for the density field ($j_1$ is the spherical Bessel function of first order). We set $V$ so that $\sigma_8$ is tied to observations when $r=8\Mpch$.

We assume our bias values are valid in the linear regime (as our analysis suggests $b_Q$ is scale-independent over at least $0.055$--$22\Mpch$), and adopt a form for the (adiabatic, CDM) linear power spectrum of $P(k) = T^2(k)k^n$ with $n=1$ (the Harrison-Zeldovich-Peebles scale-invariant case). We adopt the transfer function, $T(k)=T_0(q)$ given by Equation~29 of \citet{Eis98}, who, following \citeauthor{Bar86} (\citeyear{Bar86}; see also \citealt{Efs92}), showed that the transfer function can be characterized by its shape ($\Gamma$) via

\begin{equation}
q = \frac{k}{\Mpch}\Theta_{2.7}^2/\Gamma
\end{equation}

\noindent for a CMB temperature parameterized as $2.7\Theta_{2.7}$K. In pure CDM, $\Gamma=\Omega_m h$; however, baryons affect the power spectrum shape, and $\Gamma\rightarrow\Gamma_{\rm eff}$ (e.g., \citealt{Sug95}). Note that, as we analyze scales far smaller than the sound horizon, $\Gamma_{\rm eff}$ can be derived from the baryon fraction via $\Gamma_{\rm eff}=\alpha_r\Omega_m h$ with $\alpha_r$ given by Equation~31 of \citet{Eis98}.

Throughout this paper, we have used a concordance cosmological model with $\sigma_8=0.9$ and $\Gamma=0.21$, as quasar bias is not greatly affected by complementarily altering $\sigma_8$ and $\Gamma$. However, the conversion from $b_Q$ to $M_{DMH}$ is somewhat dependent on $\sigma_8$ and $\Gamma$. As such, we also now analyze our results in the context of a cosmological model with ($\Omega_m$, $\Omega_\Lambda$, $\sigma_8$, $\Gamma$, $h$)$ = (0.28,0.72,0.8,0.15,0.7)$, motivated by recent supernovae, large-scale structure, and CMB measurements (e.g., \citealt{Rie04,Col05,Spe06}). Note that Equation~31 of \citet{Eis98} suggests that $\Gamma=\alpha_r\Omega_m h=0.15$ is close to adopting a (realistic) baryon fraction of $\Omega_b/\Omega_m=0.185$ (e.g., \citealt{Col05}).

In Table~\ref{table:1} we display our derived values for the mass of the dark matter halos hosting quasars. The values for $M_{DMH}$ in our highest (lowest) redshift bins have been calculated based on raising (lowering) $b_Q$ by 10\% (see Equation~\ref{eqn:reweight}). A Spearman rank test shows no significant correlation between redshift and $M_{DMH}$; however such a test is not compelling for only 5 redshift bins, particularly as we don't necessarily trust our $b_Q$ corrections at low redshift (as these corrections are dependent on the less precise values of $b_Q$ derived in our highest redshift bin). We will therefore assume, as detected in Cro05, that $M_{DMH}$ is constant with redshift. Across all our individual redshift bins we obtain a weighted mean of $M_{DMH}=4.8\pm0.5\times10^{12}~{\rm h^{-1}}M_\sun$. If we instead only consider our ``best'' bins, in the range $1.0 < z < 2.1$, we obtain $M_{DMH}=5.2\pm0.6\times10^{12}~{\rm h^{-1}}M_\sun$. This value of $M_{DMH}$ deviates $\sim$1.3$\sigma$ from the value of $M_{DMH}=3.0\pm1.6\times10^{12}~{\rm h^{-1}}M_\sun$ obtained by Cro05 using a similar cosmology, and is slightly below the determination of $M_{DMH}\sim10^{13}M_\sun$ from PMN04 (see also \citealt{Por06}).

\subsection{Luminosity-Dependent Quasar Bias}

Although, in section~\ref{sec:lumclus}, we detected no significant luminosity dependence to quasar bias in any redshift bin, our detection of $1.5\sigma$ in our highest redshift bin ($\bar{z}=1.92$) is close to being significant. We can estimate the factor by which our data sample would have to increase in size before luminosity-dependent bias could be detected using our methodology.  Assuming that the noise reduction scales as the square root of the sample size, a $2\sigma$ detection in our $\bar{z}=1.92$ bin would require a sample 3.8 times larger (including stellar contamination) or 1.8 times larger (ignoring stellar contamination). Similarly, a $3\sigma$ detection would require a sample 8.6 or 4.0 times larger, respectively.

A sample size twice as large as that used in this paper should be achievable in the near future. The necessary sample size to detect luminosity-dependent biasing will be further reduced by improved photometric techniques. For example, although we attempt to restrict the range of photometric redshift over which we analyze bivariate quasar clustering to reduce the effect of catastrophic failures on our $b_Q$ estimates, some quasars in our luminosity analysis will still be placed in entirely the wrong bin of $M_g$, diluting the significance of any comparisons we make between $M_g$ bins.  Finally, we note that it is highly unlikely that luminosity-dependent quasar bias can ever be detected, via our angular analysis, to magnitudes of $g < 21$ at redshifts $z < 1.6$. However, our $1\sigma$ errors suggest that at $z < 1.6$ quasar bias changes with luminosity by less than $\pm0.6$ ($\pm0.3$ if we ignore stellar contamination).

\begin{deluxetable}{ccccc}
\tabletypesize{\scriptsize}
\tablecaption{Bivariate Estimates of the QSO bias, $b_Q$, as a function of photometric redshift $\zp$ and absolute magnitude $M_g$\label{table:3}}
\tablecolumns{5}
\tablewidth{0pt}
\tablehead{
 \colhead{} & \colhead{} & \colhead{1 parameter model} & \multicolumn{2}{c}{2 parameter model\tablenotemark{a}} \\ \cline{4-5} \\
 \colhead{$\bzp$} & \colhead{$\bar{M_g}$} & \colhead{$b_Q$} & \colhead{$a$} & \colhead{$b_Q$}
}

\startdata
\sidehead{$0.4 \leq \zp < 1.2$}
$0.85$ & -23.99 & $2.03\pm^{0.32}_{0.39}$ & $0.877\pm^{0.102}_{0.049}$ & $1.89\pm^{0.66}_{1.37}$ \\
$0.85$ & -22.82 & $2.00\pm^{0.33}_{0.39}$ & $1.000\pm^{0.000}_{0.116}$ & $2.00\pm^{0.55}_{0.64}$ \\
$0.85$ & -21.69 & $2.08\pm^{0.31}_{0.37}$ & $0.999\pm^{0.001}_{0.141}$ & $2.08\pm^{0.54}_{0.65}$ \\
\tableline
\sidehead{$1.2 \leq \zp < 1.65$}
$1.44$ & -24.84 & $2.86\pm^{0.29}_{0.32}$ & $1.000\pm^{0.000}_{0.103}$ & $2.86\pm^{0.53}_{0.50}$ \\
$1.44$ & -23.81 & $2.62\pm^{0.31}_{0.34}$ & $0.920\pm^{0.080}_{0.072}$ & $2.73\pm^{0.52}_{0.67}$ \\
$1.44$ & -23.25 & $2.68\pm^{0.30}_{0.35}$ & $0.878\pm^{0.122}_{0.064}$ & $2.87\pm^{0.55}_{0.69}$ \\
\tableline
\sidehead{$1.65 \leq \zp < 2.3$}
$1.92$ & -25.42 & $2.77\pm^{0.32}_{0.37}$ & $0.865\pm^{0.109}_{0.053}$ & $2.87\pm^{0.61}_{0.84}$ \\
$1.92$ & -24.46 & $2.25\pm^{0.45}_{0.57}$ & $0.947\pm^{0.053}_{0.089}$ & $2.36\pm^{0.76}_{1.05}$ \\
$1.92$ & -23.92 & $3.18\pm^{0.29}_{0.31}$ & $0.869\pm^{0.113}_{0.054}$ & $3.46\pm^{0.52}_{0.61}$ \\
\enddata

\tablenotetext{a}{With stellar contamination $1-a$.}
\end{deluxetable}

\section{Summary and Future Prospects}
\label{sec:summ}

In this paper, we used a sample of $\sim$300,000 photometrically classified quasars drawn from SDSS DR4 to study the evolution of quasar clustering. Our main results are:

\renewcommand{\labelenumi}{(\alph{enumi})}

\begin{enumerate}

\item Over scales of 0.16$'$ to 100$'$ ($\sim$55$\kpch$ to $35\Mpch$ at our sample's mean redshift of $z\sim1.4$) quasar clustering is well-described by Smi03 fits to dark matter clustering in simulations, a \LCDM cosmology and a single quasar bias parameter.  Quasar biasing appears to be scale-independent over this range.

\item For $\sigma_8=0.9$ and $\Gamma=0.21$ the required quasar bias is $b_Q^{\bar{z}=1.40} = 2.41\pm0.09$, rising to $b_Q^{\bar{z}=1.40} = 2.57\pm0.10$ for $\sigma_8=0.8$ and $\Gamma=0.15$.

\item Our sample alone is sufficient to rule out a constant quasar bias over $0.75 < \bar{z} < 2.28$ (for scales of $\sim$55$\kpch$ to $22\Mpch$) at, conservatively, $>99.6$\%. This significance rises to $>99.9999$\% if stellar contamination is not explicitly fit, which is likely closer to the true significance of our detection (see section~\ref{sec:fullcorr}). At $z\sim2.3$ we find $b_Q\sim3$. Considering complementary independent constraints on redshift evolution of $99.8$\% (Cro05) and $3.6\sigma$ (PMN04) from the 2QZ, it is certain that quasar bias is therefore evolving with cosmic time.

\item Using our best photometric redshift ranges, $\sigma_8$=0.8, and $\Gamma$ = 0.15, we find a mean mass for the dark matter halos hosting UVX quasars of $M_{DMH}=5.2\pm0.6\times10^{12}~{\rm h^{-1}}M_\sun$, approximately halfway between the values of $M_{DMH}$ determined from the 2QZ by PMN04 and Cro05.

\item We find no significant luminosity dependence to quasar clustering, but our analysis hints at a small dependence ($1.5\sigma$) at high redshift ($\bar{z}=1.92$). This suggests that, with improved photometric classification efficiency, a sample size of as little as 1.8 times larger ($\sim$550,000 objects) may be sufficient to detect luminosity dependence in quasar clustering at $z\sim2$. This might distinguish ``light bulb'' accretion (where quasars are either ``off'' or accrete at one efficiency; see, e.g., \citealt{Val01}), from models that allow a range of accretion efficiencies (e.g., \citealt{Lid06}).

\item Our work agrees excellently with local results from \citet{Ser06}, who studied the environments of quasars at $z < 0.4$ in DR3 (see their Figure~2). We concur that there is little luminosity dependence to quasar clustering on proper scales of $\geqsim~50\kpch$ (their $100{~h_{70}^{-1}~{\rm kpc}}$ comoving), and, also, that any weak luminosity trend is only expressed for brighter quasars ($M_g~\leqsim -24$; note, for comparison with \citealt{Ser06}, that $g-i\sim0.3$ for UVX quasars).

\item It is highly unlikely that our technique will constrain any luminosity dependence to quasar clustering to magnitudes of $g < 21$ at redshifts $z < 1.6$.  The errors on our measurements suggest that quasar bias $b_Q$ is constant at $z < 1.6$ to $\leqsim\pm0.6$.

\end{enumerate}

Although our analyses in section~\ref{sec:lumclus} uncovered no significant pattern, they were close to favoring particular trends. In the near future, the prospects for reassessing these results are excellent. Analysis of the angular clustering of photometrically classified quasars will improve not only as photometric surveys widen and deepen, increasing total numbers of objects, but also as classification efficiency improves, and is expanded to non-UVX Active Galactic Nuclei and to higher redshift. Thus, we expect the statistical power of clustering analyses of photometrically classified quasars to rapidly improve.  Further, estimates of the evolution of quasar clustering will be enhanced as quasar photometric redshift estimates tighten and catastrophic estimates diminish (e.g., \citealt{Bal07}). In combination we expect these factors to eventually allow significant constraints on the luminosity evolution of quasar clustering, particularly at $z~\geqsim~1.6$.

\acknowledgments

ADM and RJB wish to acknowledge support from NASA through grants NAG5-12578 and NNG06GH15G. GTR was supported in part by a Gordon and Betty Moore Fellowship in Data Intensive Science. RCN acknowledges the EU Marie Curie Excellence Chair for support during this work. DPS acknowledges support through NSF grant AST-0607634. The authors made extensive use of the storage and computing facilities at the National Center for Supercomputing Applications and thank the technical staff for their assistance in enabling this work.

Funding for the SDSS and SDSS-II has been provided by the Alfred P. Sloan Foundation, the Participating Institutions, the National Science Foundation, the U.S. Department of Energy, the National Aeronautics and Space Administration, the Japanese Monbukagakusho, the Max Planck Society, and the Higher Education Funding Council for England. The SDSS Web Site is http://www.sdss.org/.

The SDSS is managed by the Astrophysical Research Consortium for the Participating Institutions. The Participating Institutions are the American Museum of Natural History, Astrophysical Institute Potsdam, University of Basel, Cambridge University, Case Western Reserve University, University of Chicago, Drexel University, Fermilab, the Institute for Advanced Study, the Japan Participation Group, Johns Hopkins University, the Joint Institute for Nuclear Astrophysics, the Kavli Institute for Particle Astrophysics and Cosmology, the Korean Scientist Group, the Chinese Academy of Sciences (LAMOST), Los Alamos National Laboratory, the Max-Planck-Institute for Astronomy (MPIA), the Max-Planck-Institute for Astrophysics (MPA), New Mexico State University, Ohio State University, University of Pittsburgh, University of Portsmouth, Princeton University, the United States Naval Observatory, and the University of Washington.

\appendix

\section*{APPENDIX A. Background Methodology}

\section{Correlation Function and Model Fitting}
\label{sec:corr}

We estimate the two-point angular correlation function ($\omega$) via \citep{Lan93}

\begin{equation}
\omega(\theta) = \frac{QQ(\theta) - 2QR(\theta)}{RR(\theta)} + 1
\label{eqn:LScorr}
\end{equation}

\noindent from counts of quasar-quasar ($QQ$), quasar-random ($QR$) and random-random ($RR$) pairs. We use a random catalog 100 times larger than the data catalog. The random catalog is constructed using masks from the SDSS DR4 Catalog Archive Server and cutting both the KDE data and the random catalog to the SDSS DR4 theoretical footprint (which discards $\sim$1.7\% of the data). Our approach is detailed in Mye06, where we also discuss how points in our random catalog are assigned values of seeing and Galactic absorption. 

We estimate errors and covariance matrices using inverse-variance-weighted jackknife resampling \citep{Scr02,Zeh02,Mye05}. The jackknife method is to divide the data into $N$ pixels, then create $N$ subsamples by neglecting each pixel in turn. Note that if we {\it considered} the contribution of each pixel, rather than {\it neglecting} its contribution, this would be an inverse-variance-weighted {\it pixel-to-pixel} (also called {\it field-to-field} or {\it subsampled}) error estimate (e.g., \citealt{Mye03}). We will generally refer to the chosen length for each side of the pixels as the {\it jackknife resolution}. If we denote subsamples by the subscript $L$ and recalculate $\omega_L$ in each jackknife realization via Equation~\ref{eqn:LScorr}, then the inverse-variance-weighted covariance matrix ($C_{ij}$) can be generated as

\begin{equation}
C_{ij}=C(\theta_i,\theta_j) =\sum_{L=1}^{N}\sqrt{\frac{RR_{L}(\theta_i)}{RR(\theta_i)}}\left[\omega_{L}(\theta_i)- \omega(\theta_i)\right]\sqrt{\frac{RR_{L}(\theta_j)}{RR(\theta_j)}}\left[\omega_{L}(\theta_j)- \omega(\theta_j)\right]
\label{eqn:CM}
\end{equation}

\noindent where, $\omega$ denotes the correlation function for all data and $\omega_L$ denotes the correlation function for subsample $L$. Jackknife errors $\sigma_i$ are obtained from the diagonal elements ($\sigma_i^2 = C_{ii}$), and the normalized covariance matrix, also known as the regression matrix, is

\begin{equation}
|C| = \frac{C_{ij}}{\sigma_i\sigma_j}
\label{eqn:NCM}
\end{equation}

\noindent  The $RR_L/RR$ terms in Equation~\ref{eqn:CM} \citep{Mye05} weight by the different numbers of objects expected, due to holes, poor seeing, pixels that extend beyond the survey boundary, etc. We then estimate $\chi^2$ fits to model angular autocorrelation functions ($\omega_m$) using the inverse of the covariance matrix, and determine errors on fits from $\Delta\chi^2$, where

\begin{equation}
\chi^2 = \sum_{i,j}\left[\omega(\theta_i)-\omega_m(\theta_i)\right]C_{ij}^{-1}\left[\omega(\theta_j)-\omega_m(\theta_j)\right]
\label{eqn:chi}
\end{equation}

\noindent The simplest models we fit are power-laws of the form $\omega_m(\theta) = A\theta^{-\delta}$, where the units of $A$ (as we fit it) are ${\rm arcmin}^{\delta}$. In general we fit the more physical models discussed in Appendix~\ref{sec:modproj}.

\section{Modeling projected quasar clustering}
\label{sec:modproj}

Since the seminal scaling relations of \citet{Ham91}, many authors (e.g., \citealt{Pea94,Jai95,Pea96}) have worked to obtain precise analytical descriptions of the clustering of dark matter particles. \citeauthor{Smi03} (\citeyear{Smi03}; henceforth Smi03) married traditional approaches with a simple halo model (e.g., \citealt{Coo02}) to obtain fitting formulae that better approximate the non-linear clustering behavior of simulated CDM. The Smi03 formulae reproduce clustering in dark matter simulations to better than 3\% at (redshift) $z < 3$ for (wavenumber) $k  < 10 {~h~{\rm Mpc}^{-1}}$.

The models of Smi03 directly predict the non-linear, dimensionless power spectrum of dark matter \Pdel$(k,z)$ for a wide range of CDM cosmologies. The clustering of objects that formed in the rare peaks of a Gaussian random field is expected to be biased relative to underlying dark matter (e.g., \citealt{Kai84,Bar86}), in a manner that could depend on both scale and formation history. Thus the clustering of quasars relative to dark matter might be modeled as \Qdel$(k,z) = b_Q^2(k,z)$\Pdel$(k,z)$, where $b_Q$ is the quasar bias.

In this paper, we measure angular autocorrelations. For small angles ($\theta \ll 1~{\rm radian}$), Limber's equation can be used to project the power spectrum into the angular autocorrelation \citep{Lim53,Pee80,Pea91,Bau93} via

\begin{equation}
\omega(\theta) = \pi\int^{z=\infty}_{z=0}\int^{k=\infty}_{k=0}\frac{\Delta^2(k,z)}{k}J_0[k\theta\chi(z)]\left(\frac{{\rm d}N}{{\rm d}z}\right)^2\left(\frac{{\rm d}z}{{\rm d}\chi}\right)F(\chi)\frac{{\rm d}k}{k}{\rm d}z
\label{eqn:preprojmod}
\end{equation}

\noindent where $J_0$ is the zeroth-order Bessel function of the first kind, $\chi$ is the radial comoving distance, ${\rm d}N/{\rm d}z$ is the redshift selection function (normalized so that $\int^{\infty}_0[{\rm d}N/{\rm d}z]{\rm d}z = 1$), and ${\rm d}z/{\rm d}\chi=H_z/c=H_0\sqrt{\Omega_m(1+z)^3+\Omega_\Lambda}/c$. Strictly, $\chi$ should be the angular, or transverse, comoving distance; however, in a flat cosmology, radial and transverse comoving distances are equivalent, and the curvature term vanishes---$F(\chi)=1$. We ultimately, model the angular quasar autocorrelation function as

\begin{equation}
\omega_{QQ}(\theta) = \frac{H_0\pi}{c}\int^{\infty}_{0}\int^{\infty}_{0}b_Q^2(k,z)\frac{\Delta_{\rm NL}^2(k,z)}{k}J_0[k\theta\chi(z)]\left(\frac{{\rm d}N}{{\rm d}z}\right)^2\sqrt{\Omega_m(1+z)^3+\Omega_\Lambda}\frac{{\rm d}k}{k}{\rm d}z
\label{eqn:projmod}
\end{equation}

\noindent In theory, with sufficient data, $b_Q(k,z)$ may be directly constrained by $\omega_{QQ}$, although we will generally set $b_Q$ to be a constant and constrain it by comparing the amplitudes of $\omega_{QQ}$ and the projected matter power spectrum as a function of redshift and scale.

\begin{figure}
\plotone{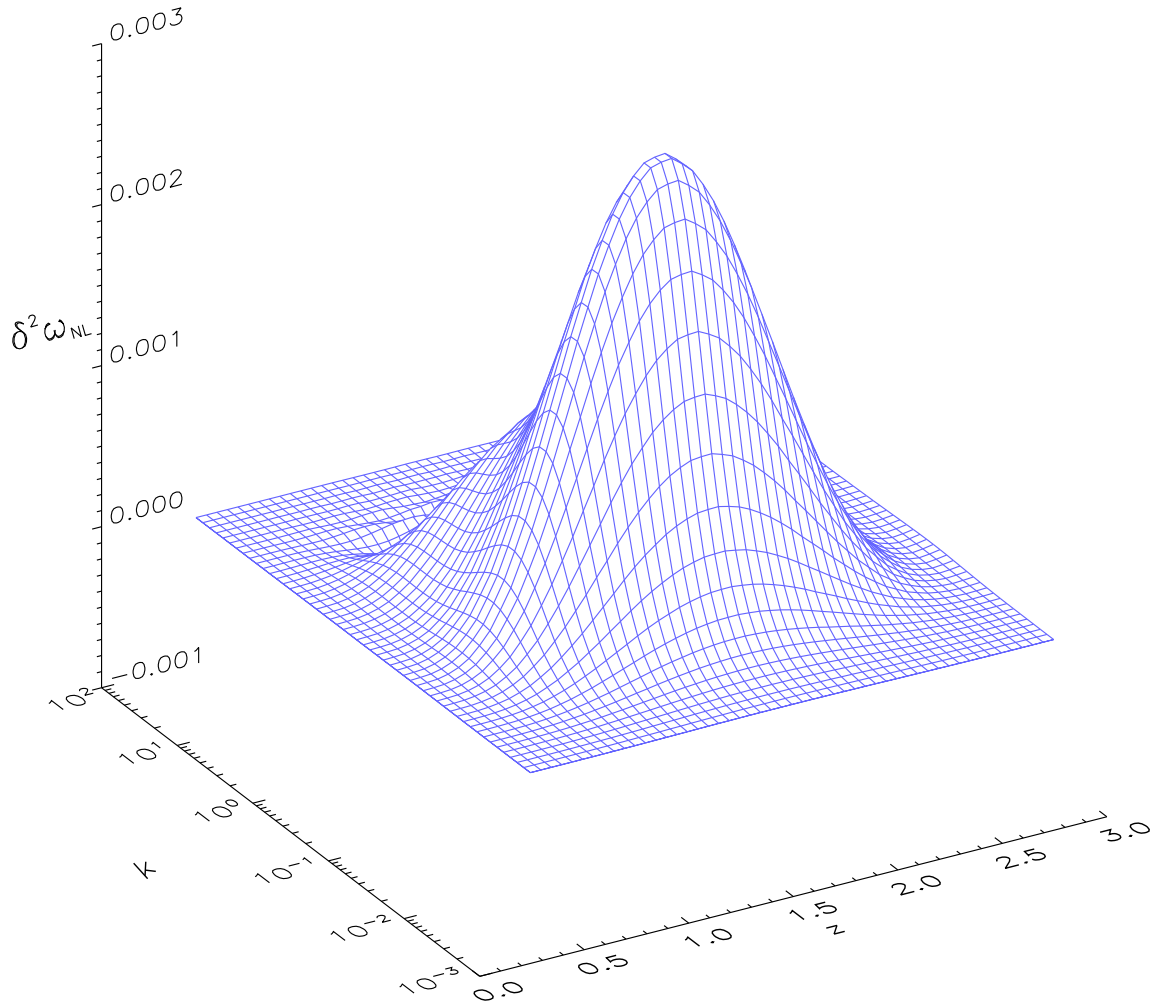}

\caption{\label{fig:surf} One example of a surface that we Monte Carlo integrate under to project the matter power spectrum into the angular autocorrelation function. The surface is plotted for $\theta=1'$, and $\delta^2\omega_{NL} = \left(\Delta_{\rm NL}^2/k^2\right)J_0[k\theta\chi(z)]\left({\rm d}N/{\rm d}z\right)^2{\rm d}z/{\rm d\chi}$ (see Equations~\ref{eqn:preprojmod} and \ref{eqn:projmod}). We obtain ${\rm d}N/{\rm d}z$ from spectroscopic matches (with DR1QSO, DR2 or the 2QZ) to our photometrically classified sample. Two individual contributions to power are apparent at 1$'$; the non-linear matter spectrum at low $k$ and a halo term at high $k$ (see Smi03).}
\end{figure}

In general, \Pdel~is not separable into individual functions of $k$ and $z$. We therefore Monte Carlo integrate under the surface described by the integrand in Equation~\ref{eqn:projmod} until the integration is evaluated to better than 1\%. To optimize this process, two points are worth noting. First, the change of variables ${\rm d}k/k = \ln(10){\rm d}\log(k)$ allows more uniform sampling in $k$-space. Second, although \Pdel~is not easily separated, we have determined that the ``parameters of the spectrum" ($k_\sigma^{-1}$, $n_{eff}$ and $C$; see Appendix~C of Smi03) can be approximated by splines to $\leqsim~0.3\%$ for at least $z < 4$ in a \LCDM cosmology. Figure~\ref{fig:surf} demonstrates a typical surface we might integrate under to obtain $\omega_{QQ}$. On scales $\leqsim~30'$ such surfaces are fairly smooth and the Monte Carlo integrations rapidly converge. Our integrations remain tractable at the 1\% level out to the largest scales we model ($\leq 100'$).

\section*{APPENDIX B. Potential Systematics}

\section{Stellar Contamination}
\label{sec:stelcon}

We have addressed sources of systematic error in some depth in Mye06. In particular, we noted that the autocorrelation of KDE objects, $\omega$, combines clustering signals from a stellar component $\omega_{SS}$ and the true quasar autocorrelation $\omega_{QQ}$.
If $a$ is the efficiency, the fraction of genuine quasars that are classified as such by the KDE technique, then 

\begin{equation}
\omega(\theta) = a^2\omega_{QQ}(\theta) + (1-a)^2\omega_{SS}(\theta) + \epsilon(\theta)
\label{eqn:stelcon}
\end{equation}

\noindent where $\epsilon$ is a tiny (theoretically zero) offset arising from $QS$, $QR$ and $SR$ cross-terms.

If the efficiency of the KDE technique is high (i.e., $a \rightarrow 1$), stellar contamination is only important as $\omega_{QQ} \rightarrow 0$. In Section~\ref{sec:fullcorr}, we fit the two-parameter model defined by Equations~\ref{eqn:projmod} and \ref{eqn:stelcon} to the DR4 KDE autocorrelation and estimate the stellar contamination ($1-a$). In doing so, we derive $a = 0.956 \pm ^{0.044}_{0.019}$, consistent with Mye06 (and with \citealt{Ric04}), and find that our best analysis of the quasar bias is therefore at angles of  $\theta~\leqsim~1^\circ$.

\section{Misclassified \Hii{} regions}

In Paper2, we discuss non-stellar misclassified objects in the KDE catalog (generally \Hii{} regions in various galaxies). The fraction of such objects in the KDE sample is too small to affect clustering measurements, except on small scales where \Hii{} regions can mimic quasar pairs, and is thus generally negligible on our scales of interest. For instance, by visually inspecting pairs of KDE objects, we estimate that on scales of 12$''$ (the smallest scale fit in this paper), the effect amounts to $\sim$1.0$\sigma$ (where $\sigma$ is the error on $\omega$), by scales of 30$''$, the effect is $<0.4\sigma$, and on larger scales, where we exceed the observed angular size of most galaxies, the effect vanishes. We are engaged in determining the regions that need masked from KDE clustering analyses because of this small effect but (unlike in Paper2 where our focus is small scales) we make no attempt to correct for the effect in this paper.

\begin{figure}
\plotone{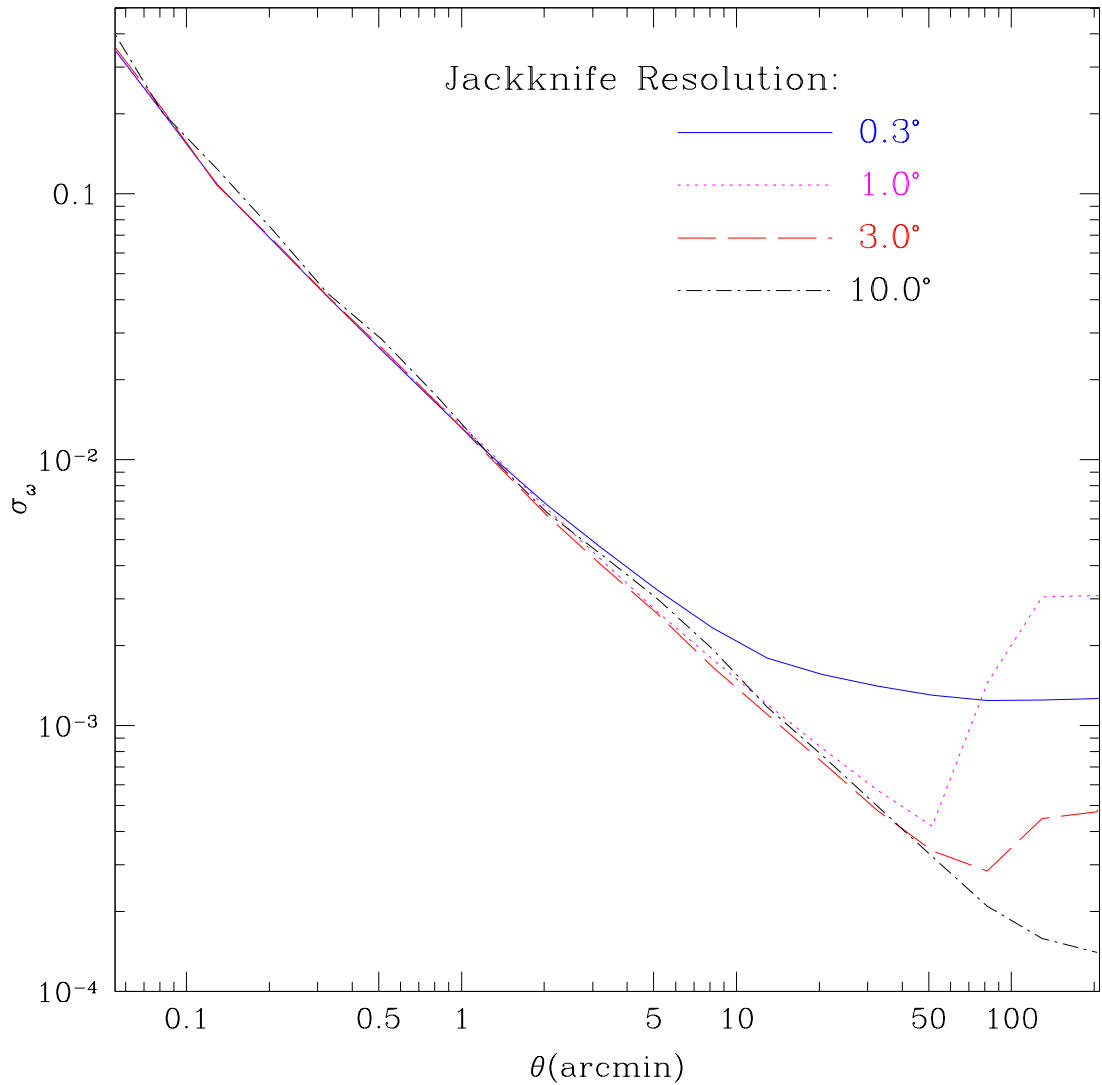}
\caption{\label{fig:errbarrescomp} Error on the correlation function, $\sigma_\omega$ at different jackknife resolutions. The error appears to be poorly defined, or flatten, around scales corresponding to the resolution. Generally, this is the same scale at which a pixel-to-pixel estimate of the error would break down. The data used to estimate $\sigma_\omega$ was the DR4 KDE sample discussed in Section~\ref{sec:data}, with an initial dust cut $A_g < 0.24$.}
\end{figure}

\begin{figure}
\plotone{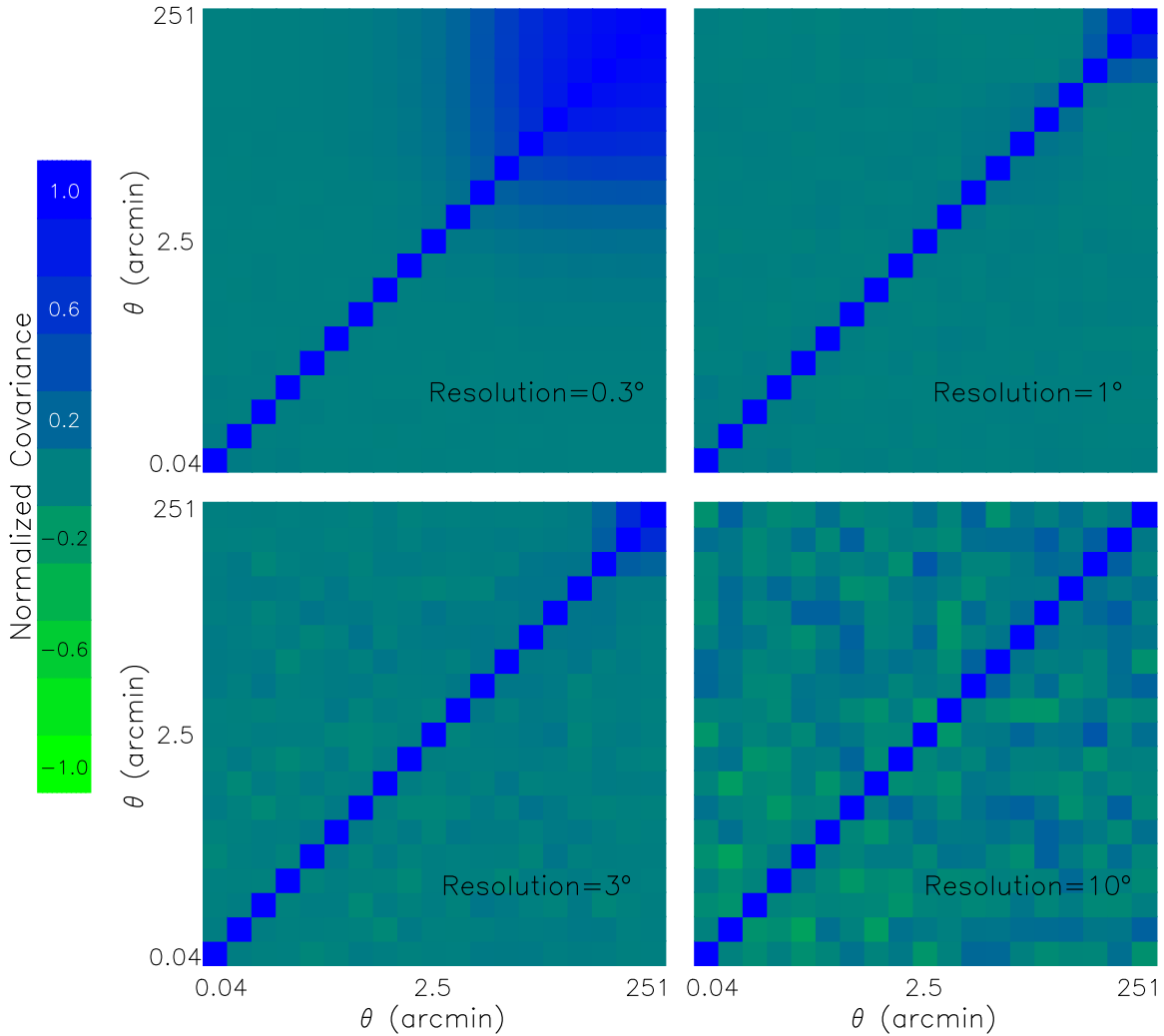}

\caption{\label{fig:covar} Normalized covariance matrix (see Equation~\ref{eqn:NCM}) at different jackknife resolutions. The covariance grows rapidly on scales larger than the jackknife resolution. On scales far smaller than the jackknife resolution, off-diagonal elements are well-mixed but still contribute to significance estimates. From top-left to bottom-right the panels represent jackknife resolutions of $0.3^\circ, 1^\circ, 3^\circ$ and $10^\circ$. The data used to estimate the covariance matrices was the DR4 KDE sample discussed in Section~\ref{sec:data}, with a nominal $A_g < 0.24$ Galactic absorption cut.}
\end{figure}

\section{Covariance and the Jackknife Resolution}

If the covariance between scales is perfectly accounted for, Equation~\ref{eqn:chi} should return identical estimates of $\chi^2$ irrespective of the jackknife resolution. Given that DR4 covers close to 7000 square degrees, we have adequate area with which to determine whether the jackknife resolution affects significance estimates. In Figures~\ref{fig:errbarrescomp} and \ref{fig:covar} we plot jackknife errors and covariance matrices for identical sets of data, obtained at different jackknife resolutions.

Figure~\ref{fig:errbarrescomp} clearly shows that error estimates are influenced by the choice of jackknife resolution, and inflate on scales similar to the resolution. However, we might expect that error discrepancies will be offset by differences in the covariance matrices plotted in Figure~\ref{fig:covar}. To test this we assume a model that, at every scale, is equal to $\omega+\sigma_\omega$ (the tip of the 1$\sigma$ upper error bar) as obtained when jackknife resampling at $10^\circ$, and calculate $\chi^2$ for other jackknife resolutions (over the 0.1$'$ to 100$'$ scales we study in this work). We find that jackknife resolutions of greater than a few degrees all return highly consistent $\chi^2$ estimates (within $\sim$2\%) but jackknife resolutions that lie increasingly within our scales of interest give increasingly inflated estimates of the significance relative to our $10^\circ$ control.

In this paper, we will use a jackknife resolution of $10^\circ$, for several reasons. The covariance matrix is well-mixed at this resolution, so that either neglecting or incorporating covariance gives similar $\chi^2$ estimates (to within 5\%). Further, at a resolution of $10^\circ$ pixel-to-pixel errors can be calculated (see Appendix~\ref{sec:corr}) for our scales of interest, and we find these agree with the jackknife estimates. Note that generalizing the ideal jackknife resolution for a given analysis is not our focus in this work; we simply suggest that when jackknifing errors, different jackknife resolutions should be tested, particularly when the resolution is similar to the scales being probed. It is tempting to ask, however; if pixel-to-pixel errors are always well-defined for resolutions that consistently recover significance estimates, why use the jackknife at all? 

\section{Additional Observational Systematics}
\label{sec:dust}

Mye06 discussed how observations that were made in poor seeing conditions or that trace absorption by dust in our Galaxy; could contaminate the true quasar clustering signal. Using a similar analysis to Mye06, with the narrower binning allowed by the larger DR4 data set, we find: (1) No clear pattern imposed by seeing variations; and (2) that an $A_g < 0.21$ cut on our DR4 KDE sample (and random catalog) removes clustering imprints caused by Galactic dust.  In agreement with Mye06, we find a clear clustering pattern in the KDE sample for $A_g~\geqsim~0.22$. We note that \citet{Yah06} found little change in the number density of KDE objects as a function of Galactic absorption, perhaps because their surface density analysis is insensitive to the {\em relative} numbers of stars and quasars assigned by the KDE technique (see expanded discussion in Mye06). We have checked that a range of cuts around $A_g~\leqsim~0.21$ (in particular, our adopted cut of $A_g < 0.18$ from Mye06) all yield statistically similar estimates of $\omega$. For our main analyses in this paper, we adopt no seeing cut and an $A_g < 0.21$ cut, which discards $\sim$12.5\% of our sampled area.

\end{document}